\documentclass[amsfonts,amsmath,prd,preprint,nofootinbib,a4paper]{revtex4}
\newcommand{\beq}{\begin{equation}}
\newcommand{\eeq}{\end{equation}}

\newcommand{\bsp}{\begin{split}}

\usepackage{epsfig,bbm,cancel,ulem}
\usepackage[breaklinks=true]{hyperref}
\usepackage{latexsym}
\usepackage[utf8]{inputenc}
\usepackage{amsmath}
\usepackage{amsfonts}
\usepackage{amssymb}
\usepackage{booktabs}
\usepackage{array}
\usepackage{tabularx}
\usepackage{multirow}
\usepackage{longtable}
\usepackage{ragged2e}
\usepackage{braket,booktabs}
\usepackage{braket,amsmath}
\usepackage{xcolor}
\usepackage{graphicx}
\usepackage{orcidlink} % for inserting orcid

\begin{document}

\title{Ambiguity in matter sector for modified gravity involving $\delta^2 \mathcal{L}_{m}/\delta g^{\mu\nu}\delta g^{\alpha\beta}$ and its implications to astrophysics and cosmology}

%\orcidlink{0000-0001-5920-8701}
\author{B. N. Jayawiguna \orcidlink{0000-0001-5920-8701}}
\email{nugrahabyon312@gmail.com}
\affiliation{Departemen Fisika, FMIPA, Universitas Indonesia, Depok, 16424, Indonesia.}

%\orcidlink{0000-0002-1493-5013}
\author{A. Sulaksono \orcidlink{0000-0002-1493-5013}}
\email{anto.sulaksono@sci.ui.ac.id}
\affiliation{Departemen Fisika, FMIPA, Universitas Indonesia, Depok, 16424, Indonesia.}

\def\changenote#1{\footnote{\bf #1}}

\begin{abstract}
Matter density ($\rho$) and radial pressure ($p$) are often used as the matter Lagrangian density ($\mathcal{L}_{m}$) because both are thermodynamically consistent and produce the same Einstein field equation (EFE) in general relativity (GR). New gravity models with explicit links between matter and geometry instead involve second-order derivatives of $\mathcal{L}_{m}$ relative to the metric tensor. So, picking either $p$ or $-\rho$ for $\mathcal{L}_{m}$ gives different effective EFEs. This confusion appears because one usually treats the four-velocity ($u_\mu$) and the metric tensor ($g_{\mu \nu}$) as independent. Here, we revisit the basics and offer a consistent framework by relaxing that assumption, thereby making the modified gravity theory independent of the choice of $\mathcal{L}_{m}$. Finally, we test this approach on neutron and quark stars (ultraviolet region) and on cosmological situations with radiation-dominated ($p=\rho/3$) equations of state (infrared region), showing how it clarifies the ambiguity in picking $\mathcal{L}_{m}$ for gravity models. 
\end{abstract}

\maketitle
\thispagestyle{empty}
%\section{Introduction}
\setcounter{page}{1}

\section{Introduction}
\label{introduction}

General Relativity (GR) is widely regarded as the most successful theory of gravity due to its conceptual simplicity and its ability to pass a broad range of tests, from ground-based experiments to observations on cosmic scales. Nevertheless, GR is not universally applicable to the study of gravity. Both theoretical considerations and observational evidence indicate that modifications to GR are necessary in the strong-gravity (ultraviolet, UV) and large-distance (infrared, IR) regimes. Further details are provided in Ref.~\cite{Shankar2022,Berti:2015itd,Blazquez-Salcedo:2022dxh}.

One of the main challenges in the infrared (IR) regime is to explain the observed late-time acceleration of the Universe \cite{WMAP:2006bqn,WMAP:2008lyn,Komatsu_2011,Li:2011sd,SupernovaCosmologyProject:1998vns,SupernovaSearchTeam:1998fmf,SupernovaCosmologyProject:2003dcn}. It is unclear whether General Relativity (GR) appears incomplete in explaining this acceleration without the cosmological constant, $\Lambda$. The cosmological constant is often interpreted as dark energy, which drives the universe's accelerated expansion. Yet, this term suffers from the well-known cosmological constant problem and coincidence problem \cite{Weinberg:1988cp}. Alternatively, the issue might be addressed by modifying the gravitational action rather than adding an extra cosmological term. A common approach is to replace the Ricci scalar $R$ in the GR action with a function $f(R)$, resulting in $f(R)$ gravity. These corrections generally involve arbitrary functions of the Ricci scalar. To explain the universe's accelerated expansion, \cite{Carroll:2003wy}, a model like $f(R)=R-\mu^4/R$ is often used, with $\mu$ a mass scale roughly equal to the present Hubble parameter: $\mu \approx 10^{-33}$ eV. In the early universe, where curvature is high, the first term dominates, and the theory reduces to Einstein GR. At late times, as curvature decreases, the Einstein term becomes less dominant, while the inverse-curvature correction becomes more significant \cite{capoziello,Kolb:1990vq,Mukhanov:1981xt,Guth:1982ec,Hawking:1982my}. The $f(R)$ framework was firmly established by the work of \cite{Starobinsky:1980te}, and has since been widely studied in cosmological and astrophysical contexts\cite{Amendola:2006we,Cooney:2009rr,Pretel:2020oae,Pretel:2020rqx,Pretel:2022plg,Alvarenga:2012bt,Bamba:2012cp,Capozziello:2007ec,Harko:2011kv,Houndjo:2011tu,Houndjo:2011fb,Jamil_2012,Nojiri:2005jg,Oikonomou:2013rba,Setare:2012vs,Shabani:2013djy,Sotiriou:2008rp}. Another significant feature of $f(R)$ gravity is its equivalence to a scalar-tensor theory. Specifically, it is a Brans-Dicke-type model ~\cite{PhysRev.124.925}, where the gravitational sector includes a nonminimal coupling to a scalar field.

The main argument concerns how ultraviolet (UV) regimes expose challenges to general relativity (GR), including black hole (BH) singularities and the existence of unusually massive horizonless compact objects in neutron star (NSs) or white dwarfs (WDs) families. Modified gravity theories introduce an extra parameter that can influence the properties of compact stars and black holes \cite{delaCruz-Dombriz:2009pzc,Cembranos:2011sr,Sheykhi:2012zz,Tang:2019qiy,Khodadi:2020cht,Khodadi:2022xtl}. Studies \cite{Yazadjiev:2014cza,Capozziello:2015yza,AparicioResco:2016xcm,Astashenok:2017dpo,Astashenok:2018iav,Feola:2019zqg} generally show mass-radius profiles for NSs increasing as the strength of modifications grows. Electron degeneracy pressure prevents gravitational collapse in WDs \cite{shapiro,lauffer,woosley}, and the Chandrasekhar limit, at approximately 1.4 solar masses \cite{Chandrasekhar:1931ftj,Chandrasekhar:1931ih}, defines their maximal mass within GR. Yet, recent studies suggest super-Chandrasekhar WDs (2.1--2.8 solar masses) exist \cite{Hicken:2007ap,Hillebrandt:2000ga,Khokhlov1993}, especially in overluminous Type Ia supernovae \cite{SNLS:2006ics,Filippenko:1992wda,Taubenberger:2007dt,Turatto:1998eq,Silverman2013,Yamanaka2009,Scalzo2010}, motivating modified gravity theories to explain this regime  \cite{Priyobarta:2026dip}. For BH, GR predicts singularities, which motivates the exploration of modified theories. GR may replace singularities with regular cores violating the strong energy condition. In $f(R)$ gravity, regular black-hole solutions arise due to higher-order curvature corrections and nonlinear electrodynamics, resulting in finite curvature invariants like the Kretschmann scalar \cite{Rodrigues:2015ayd,Rodrigues:2016fym,Hollenstein:2008hp}.

The $f(R)$ framework can be generalized by coupling curvature to the matter Lagrangian in the gravitational action. Bertolami et al.\cite{Bertolami:2007gv,Bertolami:2008ab} showed that such curvature-matter couplings create an extra force, leading to non-conservation of the matter energy-momentum tensor and possibly explaining cosmic acceleration\cite{Bisabr:2012tg}. Harko and Lobo\cite{Harko:2008qz} proposed more general theories in which the action depends on both the Ricci scalar and the matter Lagrangian, $f(R,\mathcal{L}_m)$. The resulting field equations in these models depend on the second derivative of the matter Lagrangian with respect to the metric, which must be specified. In GR, the choices $\mathcal{L}_m = p$ and $\mathcal{L}_m = -\rho$ are thermodynamically equivalent~\cite{Haghani:2023uad} and both recover the perfect fluid through action variation~\cite{Akarsu:2023lre}. However, in this class of models, these choices are not unique and lead to distinct theories with different predictions. Much literature adopts $\mathcal{L}_m = p$~\cite{Priyobarta:2026dip,Sotiriou:2008it,Pappas:2022gtt,Alam:2023grx,Asimakis:2022jel,Katirci:2013okf,Akarsu:2018zxl,Odintsov:2013iba,Harko:2020ibn,Board:2017ign,Otoniel:2025rqt}, making the second derivative vanish, as it is independent of the metric. For $\mathcal{L}_m = -\rho$~\cite{Bertolami:2008im,Mota:2024kjb}, the coupling remains nontrivial and results in different dynamics. Thus, the choice of matter Lagrangian, though equivalent in GR, introduces ambiguity in modified gravity, leading to varied predictions. The origin and implications of this ambiguity are discussed in detail in this work.

Ambiguity occurs in the variational principle because the standard fluid variables—the four-velocity $u_{\mu}$ and the metric tensor $g_{\mu\nu}$—are treated as independent degrees of freedom \cite{Akarsu:2023lre}. In this work, we address this by explicitly incorporating the normalization condition for the four-velocity into the variational procedure. We demonstrate that, under this approach, the resulting equations become independent of the specific choice of the matter Lagrangian. Our analysis is applied to several classes of modified gravity theories: $f(R, T)$, $f(R, \tau)$, $f(R, T, P)$, $f(R, TG, TGD)$, where $T\equiv g_{\mu\nu}T^{\mu\nu}$, $\tau\equiv T_{\mu\nu}T^{\mu\nu}$, $P\equiv R_{\mu\nu}T^{\mu\nu}$, $TG \equiv G_{\mu\nu}T^{\mu\nu}$, and $TGD \equiv G_{\mu\nu}\nabla^{\mu}T\nabla^{\nu}T$. 

This paper is structured as follows. First, in Section \eqref{section2}, we review the models involving curvature--matter gravity and show the second derivative of the matter Lagrangian in each model. In Section \eqref{section3}, we revisit the models using the approach introduced in Ref.~\cite{Akarsu:2023lre} and apply it to several models. We then prove that the entire formulation is independent of the choice of \(\mathcal{L}_{m}\). In Section \eqref{section4}, we transition to astrophysical applications, using the results obtained for quark-star equations of state based on the MIT bag model and neutron-star equations of state. Subsequently, in Section \eqref{section5}, we discuss the infrared-scale (cosmological) applications through the modified Friedmann and acceleration equations. Finally, we conclude our results in Section \eqref{section6}. The derivations and relevant calculations are detailed in Appendices \eqref{appA} and \eqref{appB}. In this work, we constrain ourselves to work in geometrized units, $G=c=1$, from the first stage until the astrophysical application. In the cosmological application, we follow the convention from \cite{Akarsu:2018zxl} to have a direct comparison.

\section{The modified gravity models}
\label{section2}

In this section, we show the equation of motion or EFE for several modified gravity models with $\delta^2 \mathcal{L}_{m}/\delta g^{\alpha\beta}\delta g^{\mu\nu}$ appear in their EFE. 

\subsection{$f(R,T)$ gravity}
First, we consider the linear relation between the Ricci scalar, and the scalar tensor energy-momentum. For the following details of this gravity model, see \cite{Pappas:2022gtt}. The action of the first type can compactly be written as 
\begin{equation}
S = \int d^{4}x \sqrt{-g} \left[ \frac{f(R,~T)}{16\pi} +\mathcal{L}_{m} \right],
\end{equation}
where $f(R,T)$ is a arbitrary functions of the Ricci scalar $R$ and the trace of the tensor energy-momentum $T\equiv g_{\mu\nu }T^{\mu\nu}$. The field equation, after varying the action with respect to the metric tensor, is given by
\begin{equation}
\label{efefrt}
f_{R}R_{\mu\nu} -\frac{f}{2}g_{\mu\nu} + D_{\mu\nu} f_{R} = 8\pi T_{\mu\nu}-f_{T}(T_{\mu\nu} + \Theta_{\mu\nu}),
\end{equation}
where we have used the notation 
\begin{equation}
\label{tmunu}
T_{\mu\nu} = \frac{-2}{\sqrt{-g}} \frac{\delta(\sqrt{-g} \mathcal{L}_{m})}{\delta g^{\mu\nu}} = g_{\mu\nu} \mathcal{L}_{m}-2\frac{\delta\mathcal{L}_{m}}{\delta g^{\mu\nu}}
\end{equation}
$f_{R}\equiv \partial_{R}f(R,T)$, $f_{T}\equiv \partial_{T}f(R,T)$, $D_{\mu\nu}\equiv (g_{\mu\nu}\square-\nabla_{\mu}\nabla_{\nu})$, and 
\begin{eqnarray}
\label{Theta}
\Theta_{\mu\nu} &\equiv& g^{\alpha\beta}\frac{\delta T^{\alpha\beta}}{\delta g^{\mu\nu}} \nonumber \\ &=& -2T_{\mu\nu}+ g_{\mu\nu}\mathcal{L}_{m} -2g^{\alpha\beta} \frac{\delta^2 \mathcal{L}_{m}}{\delta g^{\alpha\beta}\delta g^{\mu\nu}}.
\end{eqnarray}
Observe that the ambiguity appears in the second and the third term in left hand side of Eq. \eqref{Theta}.

\subsection{$f(R,\tau)$ gravity}
In this section, we move to the quadratic relation to the curvature terms, which was first proposed in \cite{Katirci:2013okf}. Following the latter work, it has been reported in \cite{Akarsu:2017ohj} that they elaborate the quadratic function to be a more general power form (see also the same case with a nonzero cosmological constant \cite{Board:2017ign}). For the detailed calculation, one can also refer to \cite{Akarsu:2018zxl,Alam:2023grx} approach. The action reads
\begin{equation}
S = \int d^4x \sqrt{-g} \left[ \frac{f(R,\tau)}{16\pi}+\mathcal{L}_{m}  \right],
\end{equation}
where right now the Einstein-Hilbert is modified with the additional matter term, which depends on the squared energy-momentum tensor, $f(R,\tau) = f(R) + 16\pi f(\tau)$, where $f(\tau)\equiv \alpha T_{\alpha\beta}T^{\alpha\beta}$. The field equation can effectively be expressed as follows
\begin{equation}
\label{emsgeq}
f_{R}R_{\mu\nu} - \frac{1}{2}g_{\mu\nu}f+D_{\mu\nu} f_{R} = 8\pi T_{\mu\nu} +8\pi\alpha (g_{\mu\nu}T_{\alpha\beta}T^{\alpha\beta} -2\theta_{\mu\nu}),
\end{equation}
where the new tensor $\theta_{\mu\nu}$ is defined as
\begin{eqnarray}
\label{theta}
\theta_{\mu\nu} &\equiv&\frac{\delta (T_{\alpha\beta}T^{\alpha\beta})}{\delta g^{\mu\nu}}, \nonumber \\ &=& -2\mathcal{L}_{m}\left( T_{\mu\nu}-\frac{1}{2}g_{\mu\nu}T \right) -T T_{\mu\nu}+2T_{\mu}^{~\gamma}T_{\nu\gamma}  -4 T^{\alpha\beta} \frac{\delta^2\mathcal{L}_{m}}{\delta g^{\mu\nu}\delta g^{\alpha\beta}}
\label{GammaX}
\end{eqnarray}
The field equation shows the second derivative of the matter lagrangian, which appears in the last term in left hand side of Eq. \eqref{GammaX}.

\subsection{$f(R,T,P)$ gravity}
Now, we move to another general form of the Ricci scalar and the tensor energy-momentum \cite{Haghani:2013oma,Abchouyeh:2020vfh,Odintsov:2013iba}. The action reads
\begin{equation}
S = \int d^{4}x \sqrt{-g} \left[ \frac{f(R,~T,P)}{16\pi} +\mathcal{L}_{m} \right],
\end{equation} 
the equation of motion is given by
\begin{eqnarray}
&& f_{R}R_{\mu\nu} -\frac{f}{2}g_{\mu\nu} + D_{\mu\nu} f_{R} +(T_{\mu\nu}+\Theta_{\mu\nu}) f_{T} -\nabla_{\alpha}\nabla_{(\mu}T^{\alpha}_{\nu)}f_{P} \nonumber \\ &&+ \frac{1}{2}(\Box T_{\mu\nu} f_{P} + g_{\mu\nu} \nabla_{\alpha}\nabla_{\beta}T^{\alpha\beta}f_{P}) + \Gamma_{\mu\nu}f_{P} = 8\pi T_{\mu\nu},
\end{eqnarray}
where
\begin{eqnarray}
\label{gamma}
\Gamma_{\mu\nu} = -G_{\mu\nu}\mathcal{L}_{m} -\frac{1}{2}RT_{\mu\nu} +2 R^{\alpha}_{\mu}T_{\alpha\nu}-2R^{\alpha\beta} \frac{\delta^2 \mathcal{L}_{m}}{\delta g^{\mu\nu} \delta g^{\alpha\beta}}.
\label{ThetaY}
\end{eqnarray}
We can show that, again, the ambiguity is located at the first and the fourth terms of  theleft hand side of Eq. \eqref{ThetaY}.

\subsection{$f(R,TG,TGD)$ gravity}
The last model with a longer effort is presented in \cite{Asimakis:2022jel}, which proposed a general coupling between the Einstein tensor and the matter term. We consider the actions of the form
\begin{eqnarray}
S = \int d^4x \sqrt{-g} \left[ \frac{f(R)}{16\pi} + \alpha G_{\mu\nu}T^{\mu\nu}+ \beta G_{\mu\nu}(\nabla^{\mu}T)(\nabla^{\nu}T)\right].
\end{eqnarray}
Variation of the action with respect to the metric leads
to the following field equations
\begin{equation}
f_{R}R_{\mu\nu} - \frac{1}{2}g_{\mu\nu}f+D_{\mu\nu} f_{R} = 8\pi \left[  T_{\mu\nu} + \alpha T_{\mu\nu}^{(\alpha)} + \beta T_{\mu\nu}^{(\beta)}\right],
\end{equation}
where we have defined
\begin{eqnarray}
T_{\mu\nu}^{(\alpha)} &=& g_{\mu\nu}T_{\alpha\beta}G^{\alpha\beta} + R_{\mu\nu}T - 2 G_{\nu}^{~~\alpha}T_{\mu\alpha} - 2 G_{\mu}^{~~\alpha}T_{\nu\alpha} \nonumber \\ && - R T_{\mu\nu} - \square T_{\mu\nu} + \nabla_{\alpha}\nabla_{\mu}T_{\nu}^{~~\alpha} + \nabla_{\alpha}\nabla_{\nu}T_{\mu}^{~~\alpha} \nonumber \\ && -g_{\mu\nu}(\nabla_{\alpha}\nabla_{\beta}T^{\alpha\beta})+g_{\mu\nu}\square T -\nabla_{\mu}\nabla_{\nu}T-2\Xi_{\mu\nu},
\end{eqnarray}
and
\begin{eqnarray}
T_{\mu\nu}^{(\beta)} &=& g_{\mu\nu}G^{\alpha\beta}(\nabla_{\alpha}T)(\nabla_{\beta}T) + g_{\mu\nu} R^{\alpha\beta}(\nabla_{\alpha}T)(\nabla_{\beta}T) \nonumber \\ && +R_{\mu\nu}(\nabla_{\alpha}T)(\nabla^{\alpha}T) -2 (\nabla_{\alpha}\nabla_{\nu}T)(\nabla^{\alpha}\nabla_{\mu}T)\nonumber \\ && + g_{\mu\nu}(\nabla_{\alpha}\nabla_{\beta} T)(\nabla^{\alpha}\nabla^{\beta}T) - g_{\mu\nu}(\square T)^2\nonumber \\ && -2 R_{\mu\alpha\nu\beta}(\nabla^{\alpha}T)(\nabla^{\beta} T)-2G_{\nu}^{~~\alpha}(\nabla_{\alpha}T)(\nabla_{\mu} T) \nonumber \\ && -2G_{\mu}^{~~\alpha}(\nabla_{\alpha}T)(\nabla_{\nu} T)-R (\nabla_{\mu}T)(\nabla_{\nu}T) \nonumber \\ && +2(\nabla_{\alpha}\nabla^{\alpha} T)(\nabla_{\mu}\nabla_{\nu}T)\nonumber \\ && +4G_{\alpha\beta}\nabla^{\alpha}\nabla^{\beta}T (T_{\mu\nu} + \Theta_{\mu\nu}).
\end{eqnarray}
The definition of $\Theta_{\mu\nu}$ is similar with what we have in $(\ref{Theta})$, whereas
\begin{eqnarray}
\label{Xi}
\Xi_{\mu\nu} &\equiv& G^{\alpha\beta}\frac{\delta T_{\alpha\beta}}{\delta g^{\mu\nu}}, \nonumber \\ &=& -G_{\mu\nu}\mathcal{L}_{m} +\frac{1}{2} G^{\alpha\beta}g_{\alpha\beta} (g_{\mu\nu}\mathcal{L}_{m}-T_{\mu\nu}) \nonumber \\ && -2 G^{\alpha\beta} \frac{\delta^2\mathcal{L}_{m}}{\delta g^{\mu\nu} \delta g^{\alpha\beta}}.
\end{eqnarray}
In this case, the second derivative appears in $\Theta_{\mu\nu}$ and $\Xi_{\mu\nu}$.

Referring to all models above, we will use the standard isotropic perfect fluid written below 
\begin{equation}
T_{\mu\nu} = (\rho+p)u_{\mu}u_{\nu} + p g_{\mu\nu},
\end{equation}
where $\rho,$ $p$, and $u_{\mu}$ are the density, pressure, and 4-velocity, respectively. However, in order to make the issue clearer, here we show the common procedure in the literature to calculate the second derivative, $\delta^2 \mathcal{L}_{m}/\delta g^{\alpha\beta} \delta g^{\mu\nu}$, for each choice of $\mathcal{L}_{m}$ for isotropic perfect fluid case. For $\mathcal{L}_{m}=p$ and $\mathcal{L}_{m}=-\rho$, the first derivative of $\mathcal{L}_{m}$ can be written as
\begin{eqnarray}
\label{pro}
\frac{\delta p}{\delta g^{\mu\nu}} = -\frac{1}{2}(\rho+p)u_{\mu}u_{\nu},~~~\textrm{and}~~~ \frac{\delta \rho}{\delta g^{\mu\nu}} = \frac{(\rho+p)}{2} \left(u_{\mu}u_{\nu}+g_{\mu\nu} \right),
\end{eqnarray}
The first derivative expressions above are similar to those of \cite{Haghani:2023uad,Akarsu:2023lre} by using the relation in \eqref{tmunu}. However, the second derivative for both $\mathcal{L}_{m}$ are given by
\begin{eqnarray}
\label{p2ro2}
\frac{\delta^2 p}{\delta g^{\mu\nu} \delta g^{\alpha\beta}} = 0, ~~~\textrm{and}~~~  \label{ro2} \frac{\delta^2 \rho}{\delta g^{\mu\nu} \delta g^{\alpha\beta}} =- \frac{(\rho+p)}{2} g_{\mu\alpha}g_{\nu\beta}.
\end{eqnarray}
For $\mathcal{L}_{m}=p$, the second derivative vanishes since there is no metric dependent, and it is commonly used in the literature \cite{Akarsu:2018zxl,Odintsov:2013iba,Katirci:2013okf,Board:2017ign,Harko:2011kv,Alam:2023grx,Pappas:2022gtt,Harko:2020ibn,Asimakis:2022jel}. A lot more attention is devoted to the case where $\mathcal{L}_{m}=-\rho.$ In the most literature we have explored \cite{Haghani:2013oma,Harko:2020ibn} (also see \cite{Harko:2018ayt} in section 10.1), the second derivative of the matter Lagrangian with $\mathcal{L}_{m}=-\rho$ can be dropped due to the absence of second-order or higher-order dependence on the matter Lagrangian. In \cite{Akarsu:2018zxl}, where they consider the energy momentum squared gravity (EMSG) (which is denoted by $f(R,\tau)$ in this work), the authors argue that the second derivative of $\mathcal{L}_{m}$, does not include in the \eqref{tmunu}, therefore the last term of \eqref{theta} should be vanished for both $\mathcal{L}_{m}$ leading to the same energy-momentum tensor (EMT) \eqref{theta}\footnote{For EMSG like $f(R,\tau)$, both EMT \eqref{theta} with the argument from \cite{Akarsu:2018zxl} are as follows
\begin{eqnarray}
\label{thetap}
\theta_{\mu\nu} (\mathcal{L}_{m}=p) &=& -(\rho^2 + 4\rho p+3p^2)u_{\mu}u_{\nu},\\
\label{thetaro} \theta_{\mu\nu} (\mathcal{L}_{m}=-\rho) &=& (\rho^2 -p^2)u_{\mu}u_{\nu}+(\rho^2-p^2)g_{\mu\nu},
\end{eqnarray}
where Eq. \eqref{thetap} combined with the right-hand side of \eqref{emsgeq} is similar with \cite{Akarsu:2018zxl}, while the case for Eq \eqref{thetaro} is not similar. Therefore, we can infer that these cases \eqref{thetap} and \eqref{thetaro} are not the same. }. In our opinion, according to \eqref{p2ro2}, the second derivative of the $\mathcal{L}_{m}=-\rho$ is nonzero simply because of the metric tensor $g_{\mu\nu}$ exists in the second term of \eqref{pro}, where we use the relation $\delta g_{\mu\nu}=-g_{\mu\alpha}g_{\nu\beta}\delta g^{\alpha\beta}$ in obtaining the second derivative form. The results in \eqref{pro} and \eqref{p2ro2} are obtained by using the assumption that the quantities $g_{\mu\nu}$ and $u_{\mu}$ are independent bases. Therefore, these lead to different field equations and, of course, to different cosmological and astrophysical applications. From these, we can infer that the choices are not unique since they are not the same.
\begin{eqnarray}
\Theta_{\mu\nu} (\mathcal{L}_m=p) &\neq& \Theta_{\mu\nu}(\mathcal{L}_{m}=-\rho), \\ \theta_{\mu\nu} (\mathcal{L}_m=p) &\neq& \theta_{\mu\nu}(\mathcal{L}_{m}=-\rho), \\ \Gamma_{\mu\nu}(\mathcal{L}_{m}=p) &\neq& \Gamma_{\mu\nu}(\mathcal{L}_{m}=-\rho) \\ \Xi_{\mu\nu}(\mathcal{L}_{m}=p) &\neq& \Xi_{\mu\nu}(\mathcal{L}_{m}=-\rho), 
\end{eqnarray}
while in GR, these choices of matter Lagrangian have to be the same. This paper attempts to preserve the GR profile to the curvature-matter terms, so that the universal $\mathcal{L}_{m}$ can be applied to any theory involving matter sector mixing. In the next section, we will use this assumption, using the argument used in \cite{Akarsu:2023lre}, to revise the formalism.

\section{The Revisited Model}
\label{section3}

In the previous section, the result of the second derivative presented in \eqref{p2ro2} occurs because the derivative of 4-velocity, $u_{\mu}$, in the first term was not included. In this section, we present the revised model using the formalism described in \cite{Akarsu:2023lre}. Note that the latter reference has a different starting point from that of \cite{Haghani:2023uad} in evaluating the 4-velocity, $u_{\mu}$. Therefore, we fully rely on the results from \cite{Akarsu:2023lre}. The form can be written as
\begin{equation}
\label{umu}
\frac{\delta(u_{\alpha}u_{\beta})}{\delta g^{\mu\nu}} = u_{\alpha}u_{\beta}u_{\mu}u_{\nu},
\end{equation}
where we attach the detailed derivation in the APPENDIX \eqref{appA}. Next, we will apply the form above to the modified gravity model from the previous section. The revised version is as follows. The first derivative can be written as
\begin{eqnarray}
\label{pronew}
\frac{\delta p}{\delta g^{\mu\nu}} = -\frac{1}{2}(\rho+p)u_{\mu}u_{\nu},~~~\textrm{and}~~~  \frac{\delta \rho}{\delta g^{\mu\nu}} = \frac{(\rho+p)}{2} \left(u_{\mu}u_{\nu}+g_{\mu\nu} \right),
\end{eqnarray}
where the expressions are still the same as \cite{Haghani:2023uad,Akarsu:2023lre}. The only difference is in how we compute the second derivative. In \cite{Akarsu:2023lre}, they adopt the EoS to be $p\equiv p(h,s)$ and $\rho\equiv \rho(h,s)$, where $h$ is the specific enthalpy and $s$ is the specific entropy, and evaluated at constant entropy. Therefore, the second derivative results (eqs (48) and (51) in \cite{Akarsu:2023lre}) show an adiabatic sound-speed term, $c_{s}^2$. What we want to address in this work is to make it as general as possible (without any specification of the matter properties). By incorporating the form \eqref{umu} into the first derivative, we can now obtain
\begin{eqnarray}
\label{p2new}
\frac{\delta^2 p}{\delta g^{\mu\nu} \delta g^{\alpha\beta}} &=& - \frac{(\rho+p)}{4}u_{\mu}u_{\nu}\left( g_{\alpha\beta} +2 u_{\alpha}u_{\beta} \right), \\
\label{ro2new} \frac{\delta^2 \rho}{\delta g^{\mu\nu} \delta g^{\alpha\beta}} &=& \frac{(\rho+p)}{4}\bigg(g_{\alpha\beta} g_{\mu\nu}+ g_{\alpha\beta}u_{\mu}u_{\nu} -2 g_{\mu\alpha}g_{\nu\beta} +2 u_{\alpha}u_{\beta}u_{\mu}u_{\nu} \bigg), \nonumber \\ &=&  - \frac{\delta^2 p }{\delta g^{\mu\nu}\delta g^{\alpha\beta}} + \frac{(\rho+p)}{4}g_{\alpha\beta}g_{\mu\nu}-\frac{(\rho+p)}{2} g_{\mu\alpha} g_{\nu\beta}.
\end{eqnarray}
Note that the second derivatives in \eqref{p2new} and \eqref{ro2new} are now nonzero, and we no longer have sound-speed dependence in the expression. By substituting the equations \eqref{pronew}, \eqref{p2new}, and \eqref{ro2new} above into the equations (\ref{Theta}), (\ref{theta}), (\ref{gamma}), and (\ref{Xi})\footnote{ We show the proof of all cases in the Appendix \eqref{appB}.}, we found that
\begin{eqnarray}
\label{Thetanew}
\Theta_{\mu\nu} (\mathcal{L}_m=p) &=& \Theta_{\mu\nu}(\mathcal{L}_{m}=-\rho), \\
\label{thetanew} 
\theta_{\mu\nu} (\mathcal{L}_m=p) &=& \theta_{\mu\nu}(\mathcal{L}_{m}=-\rho),\\
\label{Gammanew} 
\Gamma_{\mu\nu}(\mathcal{L}_{m}=p) &=& \Gamma_{\mu\nu}(\mathcal{L}_{m}=-\rho) \\
\label{Xinew} 
\Xi_{\mu\nu}(\mathcal{L}_{m}=p) &=& \Xi_{\mu\nu}(\mathcal{L}_{m}=-\rho).
\end{eqnarray}

By applying \eqref{umu}, we demonstrate that the field equations become structurally similar, regardless of the specific choice for $\mathcal{L}_{m}$. We compare relevant profiles to clarify how these similarities manifest in both astrophysical contexts—where conditions are extreme—and cosmological contexts—when we focus on the IR region. Our purpose is to highlight differences and commonalities across the GR case, traditional prescriptions ($\mathcal{L}_{m}=p$ and $\mathcal{L}_{m}=-\rho$), and our new prescriptions derived from Eqs. \eqref{Thetanew}-\eqref{Xinew}.

\section{Astrophysical Application: Quark Star and Neutron Star}
\label{section4}

In this section, we applied our revised model to the quark and neutron star EOS with the $f(R, T) = R + 2\chi T$ gravity model as an example and compared the field equations' behavior written in \cite{Pappas:2022gtt}. For the QS, the well-known MIT bag model can be expressed as

\begin{equation}
\rho =3p+4B.
\end{equation}
The latter describes a fluid made up of down, up, and strange quarks. The $B$ denotes a bag constant with the value $B=60 $ MeV/fm$^3$. For the NS, we use MPA1 \cite{Muther:1987xaa}. In evaluating the component, we adopt the spherically symmetric line element
\begin{equation}
ds^2 = -e^{\nu} dt^2 + e^{\lambda} dr^2 + r^2 (d\theta^2 + r^2 \sin\theta d\phi^2),
\end{equation}
where the quantities $\nu(r)$ and $\lambda(r)$ are depend solely on radial coordinate. Next, we will use the $f(R,T)$ gravity model and analyze the new prescription using \eqref{p2new} and \eqref{ro2new} and old prescription ($\mathcal{L}_{m}=p$ and $\mathcal{L}_{m}=-\rho$) using \eqref{p2ro2}.

\subsection{New Prescription}

In the gravity part, the field equation after using the new formalism \eqref{Thetanew} reads
\begin{equation}
G_{\mu\nu}=8\pi T_{\mu\nu} + \chi g_{\mu\nu} T,
\end{equation}
which is relatively simpler since $\Theta_{\mu\nu}=-T_{\mu\nu}$, so the last term in the right-hand side of \eqref{efefrt} vanishes. Hence, each component for the new formalism is given by
\begin{eqnarray}
\label{mnew}
m' &=& 4\pi r^2 \rho + \frac{\chi r^2}{2}(\rho-3p), \\ \nu' &=& \frac{2m+8\pi r^3 p+\chi r^3(3p-\rho)}{r(r-2m)},\\ \label{pnew} p'&=&-\frac{8\pi ~(\rho+p)}{8\pi + 3\chi^2} \left[\frac{m+4\pi r^3 p +\frac{\chi}{2}r^3 (3p-\rho)}{r(r-2m)}  \right] \nonumber \\ && + \frac{\chi}{8\pi+3\chi}\rho'.
\end{eqnarray}
In the limit $\chi \rightarrow 0,$ the equations above reduce to GR case.
\subsection{Old Prescription ($\mathcal{L}_{m} = p$)}

In order to see the discrepancy compared to the new prescription, we show the field equation and its components for the old one with $\mathcal{L}_{m}=p$ \cite{Pappas:2022gtt}. The equation is given by
\begin{equation}
G_{\mu\nu} = 8\pi T_{\mu\nu} + \chi g_{\mu\nu}T + 2\chi (T_{\mu\nu}-pg_{\mu\nu}),
\end{equation}
and 
\begin{eqnarray}
\label{mold}
m' &=& 4\pi r^2 \rho + \frac{\chi}{2}r^2(3\rho-p),\\ \nu' &=& \frac{2m+8\pi r^3 p+\chi r^3(3p-\rho)}{r(r-2m)}, \\ \label{pold} p' &=& -(\rho+p) \left(\frac{8\pi+2\chi}{8\pi+3\chi}\right)  \left[ \frac{m+4\pi r^3 p + \frac{\chi}{2}r^3 (3p-\rho)}{r(r-2m)}   \right]\nonumber \\ && +\frac{\chi}{8\pi+3\chi}\rho'.
\end{eqnarray}
It is worth noting that in obtaining the mass, metric, and TOV equations above, they use \eqref{pro} and \eqref{p2ro2}.
\subsection{Old Prescription ($\mathcal{L}_{m} = -\rho$)}

For $\mathcal{L}_{m}=-\rho$, the field equations are given by
\begin{equation}
G_{\mu\nu} = 8\pi T_{\mu\nu} + \chi g_{\mu\nu}T + 2\chi [T_{\mu\nu}+g_{\mu\nu}(2\rho+p)],
\end{equation}
and
\begin{eqnarray}
\label{moldminro}
m' &=& 4\pi r^2 \rho - \frac{\chi}{2}r^2(\rho+5p),\\ \nu' &=& \frac{2m+8\pi p r^3+\chi r^3(3\rho+7p)}{r(r-2m)}, \\ \label{poldminro} p' &=& -(\rho+p) \left(\frac{8\pi+2\chi}{8\pi-3\chi}\right)  \left[ \frac{m+4\pi r^3 p + \frac{\chi}{2}r^3 (3\rho+7p)}{r(r-2m)}   \right]\nonumber \\ && -\frac{3\chi}{8\pi+2\chi}\rho'.
\end{eqnarray}

\subsection{Results}

\begin{figure}[h!]
	\centering
	\includegraphics[width=0.45\textwidth]{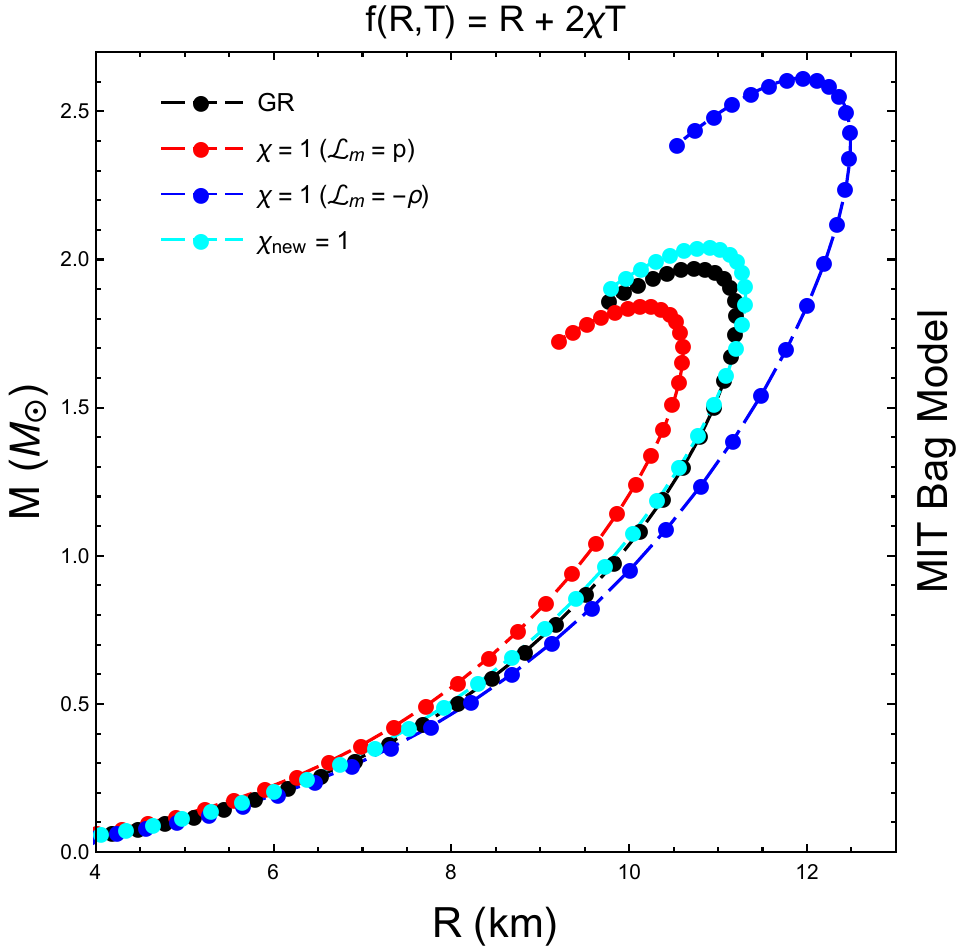}
	\includegraphics[width=0.45\textwidth]{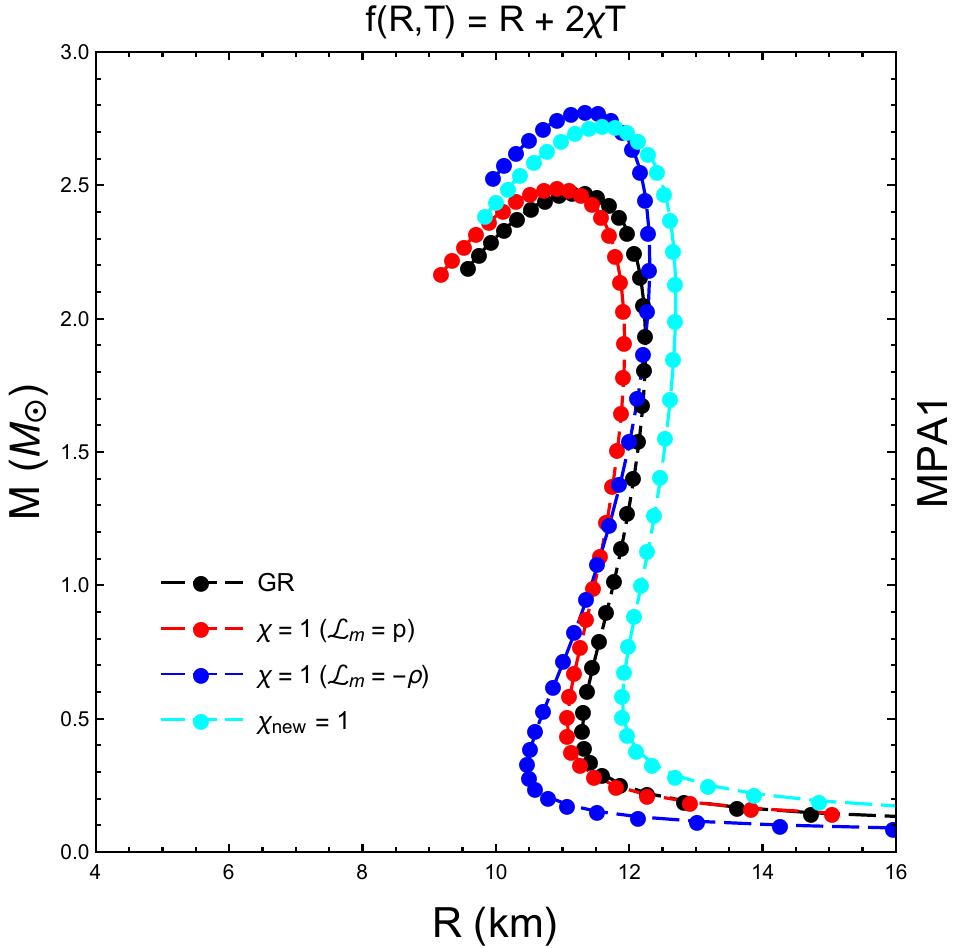}
	\includegraphics[width=0.45\textwidth]{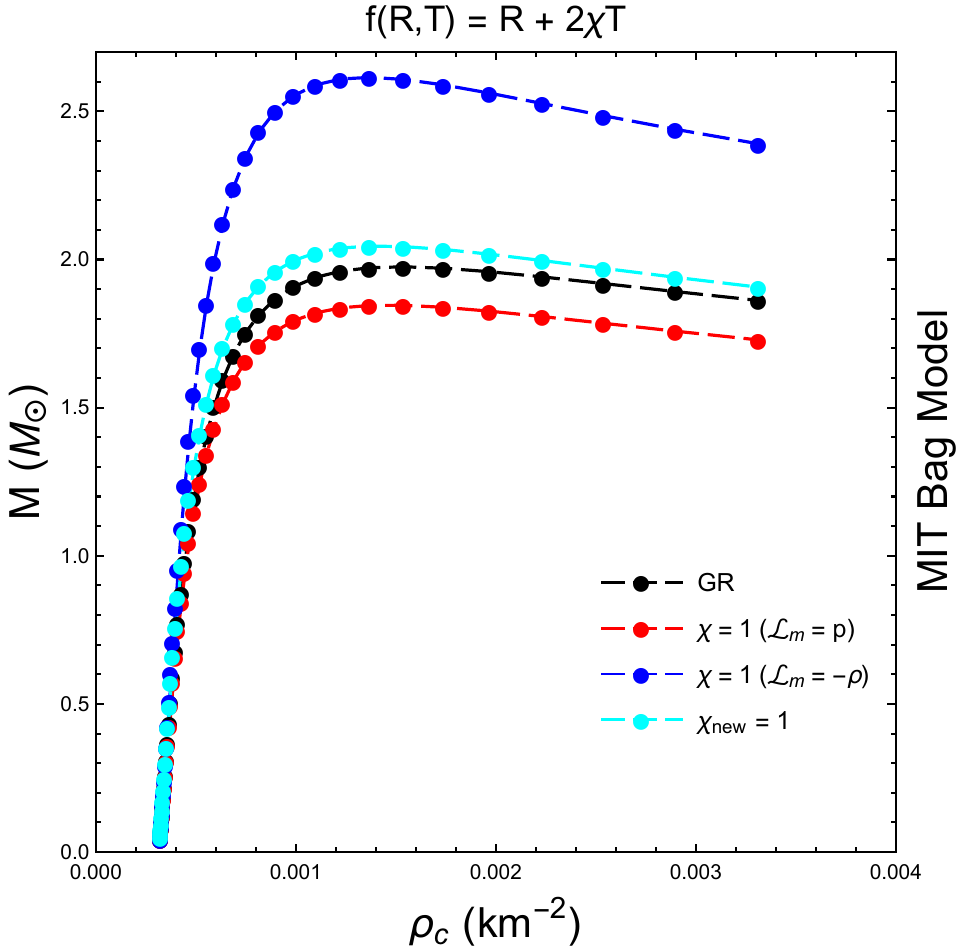}
	\includegraphics[width=0.45\textwidth]{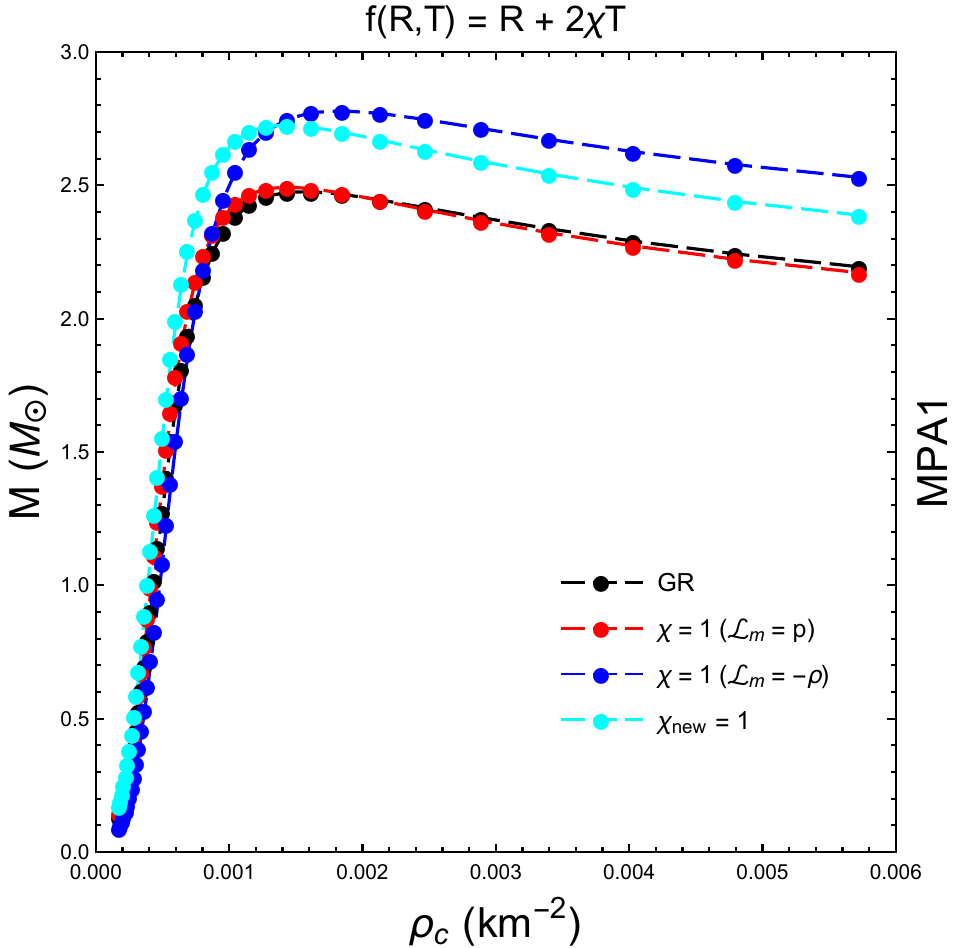}
	\caption{The mass and radius relation (top) and the mass and central density (bottom) for quark star with the MIT bag model, and the neutron star with MPA1 model.}
	\label{fig:1}
\end{figure}
The field equations of these three cases take different forms, as shown in the upper line of Fig.~\ref{fig:1}. We plot the mass-radius relation using the quark star EoS with the standard MIT bag model and a representative neutron star EoS, MPA1. The variable, $\chi$, with the subscript "new", corresponds to the dimensionless constant from the new description in (\ref{mnew})-(\ref{pnew}), while the others are from the field equation with the choices either $\mathcal{L}_{m}=p$ or $\mathcal{L}_{m}=-\rho$. The black dotted line is the profile for GR. For a representative example, if we chose $\chi=1$, we found that the old prescriptions differ, and the new prescriptions' profiles are very close to GR for quark stars with the MIT bag model and higher than GR for neutron stars with the MPA1 model. In detail, the distinct impact on the maximum mass of NS and QS originates from the different nature of their EoS. The MIT bag model for QS is approximately linear, making the compact object structure less sensitive to the $\chi$ corrections in $f(R, T)$ gravity. Consequently, the modification mainly shifts the maximum mass while the radius is nearly unaltered.

In contrast, the MPA1 EoS for NSs strongly depends on density, so the additional dimensionless constant, $\chi$, more significantly modifies the M-R profile at high central densities. This leads to a larger deviation in both the radius and maximum mass compared to QS. Therefore, we can infer that the same modified gravity correction produces qualitatively different effects on NSs and QSs due to their distinct matter properties. However, we also include supportive evidence of this explanation by displaying the mass (in solar masses) with respect to the central density of each EoS in the lower panel in Fig.~\ref{fig:1}. Both profiles are stable up until the maximum mass, and become unstable if we increase the central pressure (or density) further.

\section{Cosmological application}
\label{section5}

In this section, we apply the ambiguity choices of $\mathcal{L}_{m}$ in the IR regime. In the previous section, we saw that the choices of $\mathcal{L}_{m}$ reveal a discrepancy in the M-R diagram of horizonless compact objects, and our new formalism shows that the new profile is very close to GR. Therefore, it is intriguing to investigate the latter effect in the early stages of the universe. In this section, we use the energy-momentum squared gravity (EMSG) with $p=w\rho$, where dust ($w = 0$), radiation ($w = 1/3$), and stiff EoS ($w = -1$). Next, we show our new prescription and the old prescription ($\mathcal{L}_{m}=p$ and $\mathcal{L}_{m}=-\rho$). The assumption of the cosmological constant is zero, and we use $k=0$ in the Friedmann–Lemaître–Robertson–Walker (FLRW) metric, $ds^2 = -dt^2 + a(t)^2 d\vec{x}^2$, where $a(t)$ is the scale factor.

\subsection{GR}
In the GR case, we switched off the parameter contribution from EMSG. Therefore, the Friedmann equation, the acceleration equation, and the conservation equation can be written as
\begin{eqnarray}
3H^2 &=& 8\pi \rho, \\ -2\dot{H} -3H^2 &=& 8\pi w \rho, \\ 3H(1+w)\rho + \dot{\rho} &=& 0,
\end{eqnarray}
where the equations for the specific case of $w$ are not presented here. The analytic solutions for the dust ($w=0$) case are
\begin{eqnarray}
H = \frac{2}{3t},~~~\textrm{and}~~~ \rho = \frac{1}{6\pi t^2},
\end{eqnarray}
while for radiation ($w=1/3$) case
\begin{eqnarray}
H = \frac{1}{2t},~~~\textrm{and}~~~ \rho_r = \frac{3}{32\pi t^2}.
\end{eqnarray}
Lastly, for the vacuum energy case, the solution for the Hubble time and density is constant. Later, we will compare the EMSG model with the new and old prescriptions with the GR case.

\subsection{Old Prescription ($\mathcal{L}_{m}=p$)}
In the old prescription, we showed that the choices $p$ and $-\rho$ do not represent the same equation of motion. To present it in detail, we give the general expression with the specific value of $w$ afterward. In general $w$, equations are
\begin{eqnarray}
8\pi \left[\rho + \alpha (\rho^2 + 8 w \rho^2 + 3 w^2 \rho^2)\right] &=& 3H^2, \\ 8\pi (w\rho + \alpha \rho^2 + 3\alpha w^2 \rho^2) &=& -2\dot{H} -3H^2, \\ \left[1+2(1+8w+3w^2)\alpha \rho\right] \dot{\rho} + 3 \left[1+2\alpha (1+3w)\rho\right] (1+w) H \rho &=& 0.
\end{eqnarray}

It is worth noting that in the case where $\alpha=0$, the latter equations reduce to the GR case. 

\subsubsection{Dust ($w=0$)}
For the dust case in EMSG, the field equations are given by
\begin{eqnarray}
3H^2 &=& 8\pi (\rho+\alpha \rho^2), \\ -2\dot{H} - 3H^2 &=& 8\pi \alpha \rho^2, \\ (1+2\alpha\rho) (\dot{\rho} + 3H\rho) &=& 0.
\end{eqnarray}
In evaluating the Hubble time and the density, we can only use the first two equations. The density expression after using the first line reads
\begin{equation}
\rho = \frac{1}{2\alpha} \left( -1 + \sqrt{1+  \frac{3\alpha H^2}{2\pi} }      \right),
\end{equation}
where we choose the physical root since the latter equation will reduce to GR when $\alpha\rightarrow 0$. After some algebra, the analytic solution reads
\begin{eqnarray}
H = \frac{4\pi t}{6\pi t^2 - \alpha},~~~\textrm{and}~~~ \rho = \frac{6\pi t^2}{\alpha(\alpha-6\pi t^2)}.
\end{eqnarray}
We can see that under the influence of the EMSG, the expressions in the above equations remain relatively simple.
\subsubsection{Radiation ($w=1/3$)}

In the reference \cite{Akarsu:2018zxl} ($\mathcal{L}_m=p$), the field equations read
\begin{eqnarray}
\label{friedmannrmsg}
3H^2 &=& 8\pi \rho_r (1+4\alpha \rho_r), \\ -2\dot{H} -3H^2 &=& 8\pi \frac{\rho_r}{3} (1+4\alpha \rho_r),\\ \label{consemsg} (1+8\alpha\rho ) \dot{\rho} + 4H \rho(1+4\alpha\rho) &=& 0.
\end{eqnarray}
In \cite{Akarsu:2018zxl}, the solution for the Hubble parameter is given by $H(t)=1/(2t+C)$, where $C$ is an integration constant, and the authors choose \(C=0\). Therefore, we later follow the same convention in both the new formalism and the old prescription with \(\mathcal{L}_{m}=-\rho\). The field equations above are also similar to \cite{Akarsu:2023agp}. In the first two lines, the right-hand side has the same $(1+4\alpha\rho_r)$ factor; therefore, the Hubble time is the same as in the GR case. Hence, the density can be obtained by inserting $H$ into the field equation
\begin{eqnarray}
H = \frac{1}{2t},~~~\textrm{and}~~~\rho_r = \frac{1}{8\alpha}\left(\sqrt{1+\frac{3\alpha}{2\pi t^2}}-1\right),
\end{eqnarray}
where, again, we choose the root of the density so that they satisfy GR case in the limit $\alpha \rightarrow 0 $.

\subsubsection{Stiff ($w=-1$)}
This choices lead us to the vacuum energy case
\begin{eqnarray}
3H^2 &=& 8\pi (\rho-4\alpha \rho^2), \\ -2\dot{H} -3H^3 &=& 8\pi (-\rho +4\alpha\rho^2), \\ \dot{\rho} &=& 0.
\end{eqnarray}
From the above expressions, it is shown that $\dot{H}=0$ and $\dot{\rho}=0$.

\subsection{Old Prescription ($\mathcal{L}_{m}=-\rho$)}
In this subsection, we emphasize the field equations for $\mathcal{L}_{m}=-\rho$ for general $w$ as well as the specific case. For the general case, the equations are as follows
\begin{eqnarray}
8\pi (\rho+3\alpha\rho^2 + 14w\alpha\rho^2-9w^2 \alpha\rho^2)&=& 3H^2, \\ 8\pi (w\rho-\alpha\rho^2+4w\alpha\rho^2+9w^2\alpha\rho^2) &=& -2\dot{H} -3H^2, \\3H \rho [1+w+2\alpha(1+9w)\rho] + [1+2\alpha\rho(3+14w-9w^2)] \dot{\rho}  &=& 0.
\end{eqnarray}

It is important to note that the EMSG case in $\mathcal{L}_{m}=-\rho$ has different field equations with $\mathcal{L}_{m} = p$.

\subsubsection{Dust ($w=0$)}
In the dust case, the field equations are given by
\begin{eqnarray}
3H^2 &=& 8\pi (\rho + 3\alpha \rho^2), \\ 2 \dot{H} + 3H^2 &=& 8\pi \alpha \rho^2, \\ 3H\rho(1+2\alpha\rho) +(1+6\alpha\rho)\dot{\rho} &=& 0.
\end{eqnarray}
The solution for Hubble time and density, after some algebra, can be expressed into
\begin{eqnarray}
t= \frac{\sqrt{18 \alpha  H^2+4 \pi }-2 \sqrt{6} \sqrt{\alpha } H \tanh ^{-1}\left(\frac{2 \sqrt{3} \sqrt{\alpha } H}{\sqrt{9
			\alpha  H^2+2 \pi }}\right)+2 \sqrt{6} \sqrt{\alpha } H \tanh ^{-1}\left(\sqrt{\frac{3}{2 \pi }} \sqrt{\alpha }
	H\right)+2 \sqrt{\pi }}{6 \sqrt{\pi } H} \nonumber \\ 
\end{eqnarray}
\begin{eqnarray}
\rho = \frac{1}{6\alpha} \left( \sqrt{1+ \frac{9\alpha H^2}{2\pi}} -1 \right).
\end{eqnarray}
Note that it is complicated to transform $H(t)$, so we leave as it is and implement them numerically.

\subsubsection{Radiation ($w=1/3$)}
The field equations can be written as follows
\begin{eqnarray}
3H^2 &=& 8\pi \rho_r \left( 1+\frac{20}{3}\alpha \rho_r \right),
\\-2\dot{H}-3H^2 &=& 8\pi \frac{\rho_r}{3} (1+4\alpha \rho_r),\\ 12 H \rho (1+6\alpha\rho) + (3+40\alpha\rho) \dot{\rho} &=& 0.
\end{eqnarray}
The solutions are as follows

\begin{equation}
\rho_r = \frac{3}{40\alpha} \left( \sqrt{1+\frac{10\alpha H^2}{\pi}} -1  \right),
\end{equation}

\begin{eqnarray}
t = \frac{2 \left(\sqrt{10 \alpha  H^2+\pi }+\sqrt{\pi }\right)-11 \sqrt{\alpha } H \tanh ^{-1}\left(\frac{11 \sqrt{\alpha }
		H}{2 \sqrt{10 \alpha  H^2+\pi }}\right)+11 \sqrt{\alpha } H \tanh ^{-1}\left(\frac{9 \sqrt{\alpha } H}{2 \sqrt{\pi
	}}\right)}{8 \sqrt{\pi } \alpha  H}.
\end{eqnarray}

\subsubsection{Stiff ($w=-1$)}
In the vacuum energy case, the final results are
\begin{eqnarray}
3H^2 &=& 8\pi (\rho-20\alpha\rho^2), \\ -2\dot{H}-3H^2 &=& 8\pi (-\rho+4\alpha\rho^2),\\ 48\alpha H \rho^2 + (40\alpha \rho +1)\dot{\rho}  &=& 0.
\end{eqnarray}
In this case, the Hubble time and the density are not constant. 

\begin{eqnarray}
\rho = \frac{1}{40\alpha} \left(1- \sqrt{1-\frac{30\alpha H^2}{\pi}} \right)
\end{eqnarray}

\begin{eqnarray}
t=-\frac{15 \alpha  H^2 \left(-2 \sqrt{1-\frac{30 \alpha  H^2}{\pi }}-3\right)+\sqrt{\pi } \sqrt{\pi -30 \alpha  H^2}+\pi }{54
	\alpha  H^3}
\end{eqnarray}
In the limit $\alpha\rightarrow0$, the density reads $\rho = 3H^2/8\pi$ and $\dot{H}=0.$ Therefore, the Hubble time is constant, so does the density.

\subsection{New formalism}

In this formalism, the second-derivative relations given by Eqs.~\eqref{p2new} and \eqref{ro2new} are employed to derive the cosmological field equations. An important consequence of this approach is that the resulting equations are independent of the choice of the matter Lagrangian, namely $\mathcal{L}_{m}=p$ or $\mathcal{L}_{m}=-\rho$. This resolves the ambiguity that commonly appears in theories with explicit matter-curvature couplings, where different matter Lagrangians generally lead to inequivalent field equations. Furthermore, the present formalism yields $\theta_{\mu\nu}=0$, such that Eq.~\eqref{emsgeq} contains only two contributions on its right-hand side. Consequently, the modified Einstein equations acquire a considerably simpler structure while still incorporating the quadratic energy-density correction governed by the coupling parameter $\alpha$. The cosmological dynamics are therefore modified solely through the effective energy density and pressure, without introducing additional source terms originating from $\theta_{\mu\nu}$. Assuming a spatially flat Friedmann-Lemaître-Robertson-Walker (FLRW) spacetime filled with a perfect fluid satisfying the barotropic equation of state

\begin{equation}
p=w\rho,
\end{equation}
the trio equations can be expressed as 
\begin{eqnarray}
3H^2 &=& 8\pi [\rho-\alpha (\rho^2 + 3w^2 \rho^2)], \\ -2\dot{H} -3H^2 &=& 8\pi (w\rho + \alpha \rho^2 + 3w^2 \alpha \rho^2), \\ 3(1+w)H\rho + [1-2\alpha\rho(1+3w^2)] \dot{\rho}  &=& 0.
\end{eqnarray}
The first equation shows that the correction to the effective energy density is proportional to $\rho^2$, implying that deviations from GR become increasingly important. In the limit $\alpha \rightarrow 0$, the results will come back to GR.

\subsubsection{Dust ($w=0$)}

For dust case, the expression reads
\begin{eqnarray}
3H^2 &=& 8\pi\rho (1-\alpha \rho), \\ -2\dot{H} -3H^2 &=& 8\pi\alpha \rho^2, \\ 3H\rho + (1-2\alpha\rho)\dot{\rho}  &=& 0.
\end{eqnarray}

\begin{equation}
\rho = \frac{1}{2\alpha} \left( 1- \sqrt{1-\frac{3\alpha H^2}{2\pi}} \right)
\end{equation}

\begin{eqnarray}
t = \frac{\sqrt{4-\frac{6 \alpha  H^2}{\pi }}+\sqrt{\frac{6}{\pi }} \sqrt{\alpha } H \sin ^{-1}\left(\sqrt{\frac{3}{2 \pi }}
	\sqrt{\alpha } H\right)+2}{6 H}
\end{eqnarray}

\subsubsection{Radiation ($w=1/3$)}

For a radiation-dominated universe, the corresponding equations become
\begin{eqnarray}
\label{friedman}
3H^2 &=& 8\pi \rho_r \left( 1-\frac{4}{3}\alpha \rho_r \right),
\\-2\dot{H}-3H^2 &=& 8\pi \frac{\rho_r}{3} (1+4\alpha \rho_r), \\ \label{cons} 12H\rho + (3-8\alpha\rho) \dot{\rho}&=& 0
\end{eqnarray}

Compared with the dust case, the quadratic correction is enhanced by the radiation equation of state. Since the radiation era corresponds to much higher energy densities, the nonlinear contribution proportional to $\rho_r^2$ is expected to play a more prominent role in studying the relationship between Hubble time, radiation density, and cosmic time.

\begin{equation}
\rho_r = \frac{3}{8\alpha} \left(  1- \sqrt{1-\frac{2\alpha H^2}{\pi}}  \right)
\end{equation}

\begin{eqnarray}
t = \frac{\sqrt{\pi -2 \alpha  H^2}+\sqrt{2} \sqrt{\alpha } H \sin ^{-1}\left(\sqrt{\frac{2}{\pi }} \sqrt{\alpha }
	H\right)+\sqrt{\pi }}{4 \sqrt{\pi } H}
\end{eqnarray}

\subsubsection{Stiff ($w=-1$)}

For the vacuum-energy equation of state, the field equations become
\begin{eqnarray}
3H^2 &=& 8\pi (\rho-4\alpha\rho^2), \\    -2\dot{H} - 3H^2 &=& 8\pi (-\rho + 4\alpha \rho^2), \\
\dot{\rho}(1-8\alpha\rho) &=&0
\end{eqnarray}

In the conservation equation, we can infer that the density is still constant; it can be either $\rho=\textrm{constant} ~(H=\textrm{constant})$, or $\rho=1/8\alpha$ ($H=\sqrt{\pi/6\alpha}$).

\subsection{Results}

\begin{figure}[h!]
	\centering
	\includegraphics[width=0.45\textwidth]{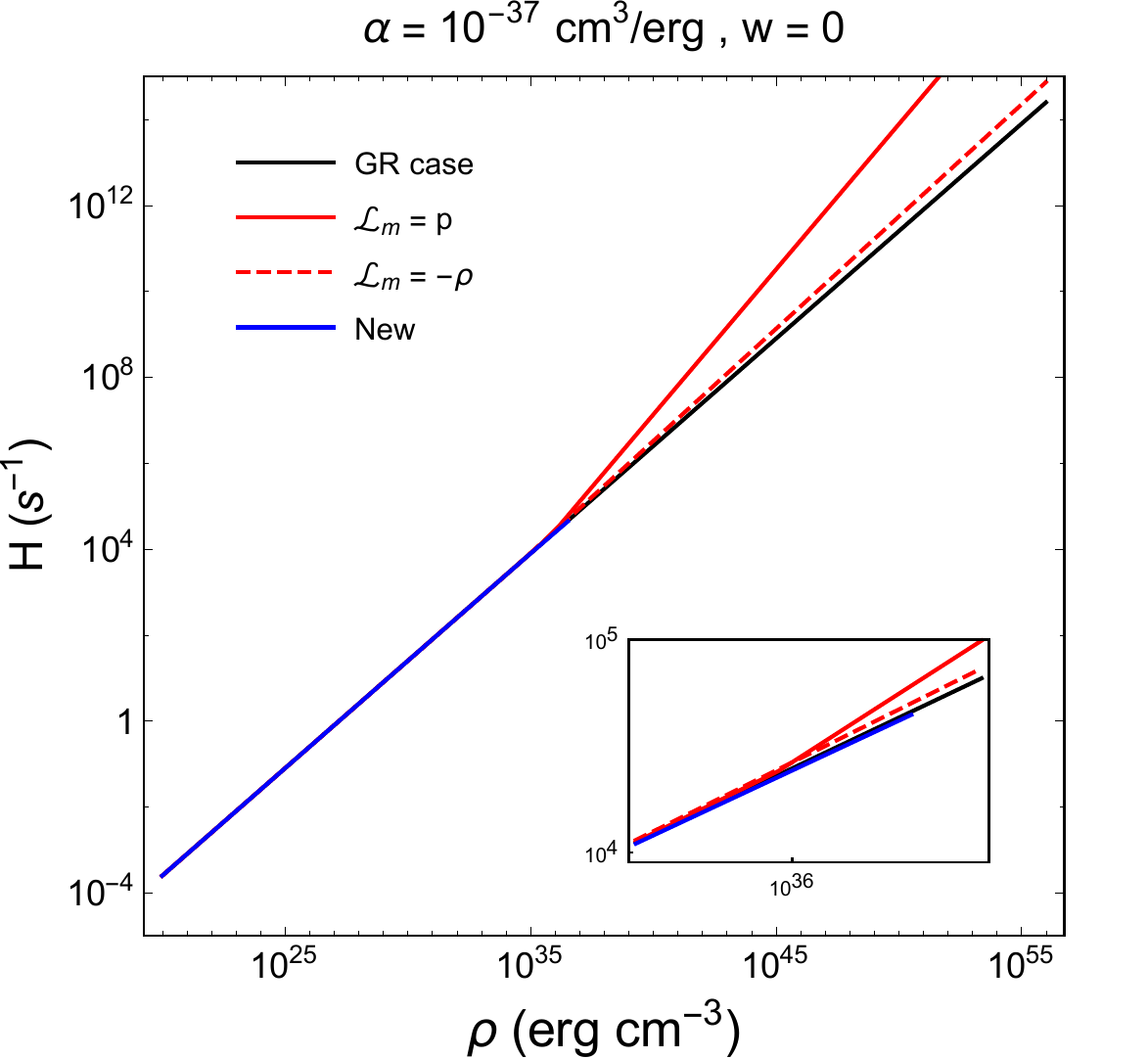}
	\includegraphics[width=0.45\textwidth]{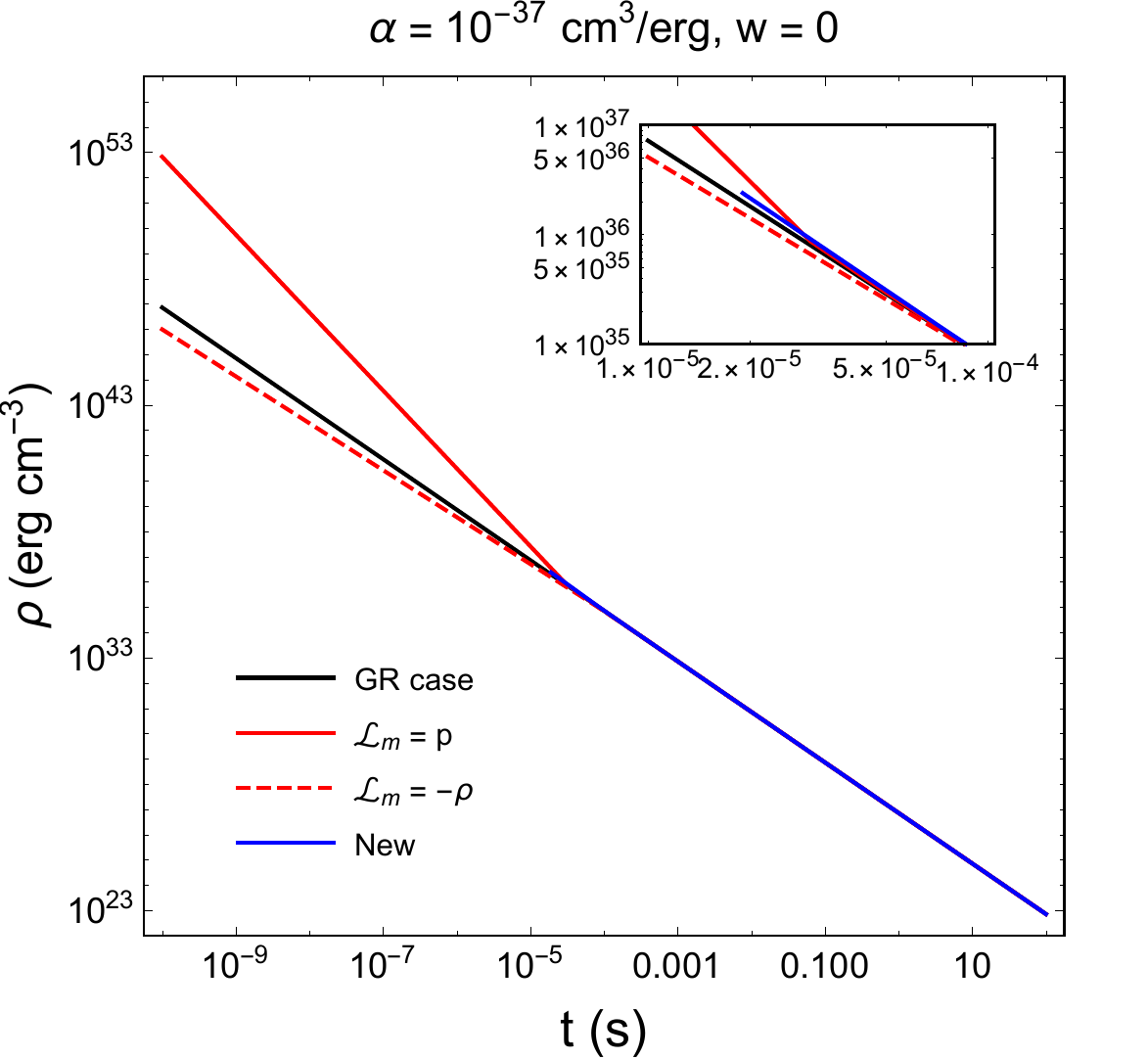}
	\includegraphics[width=0.45\textwidth]{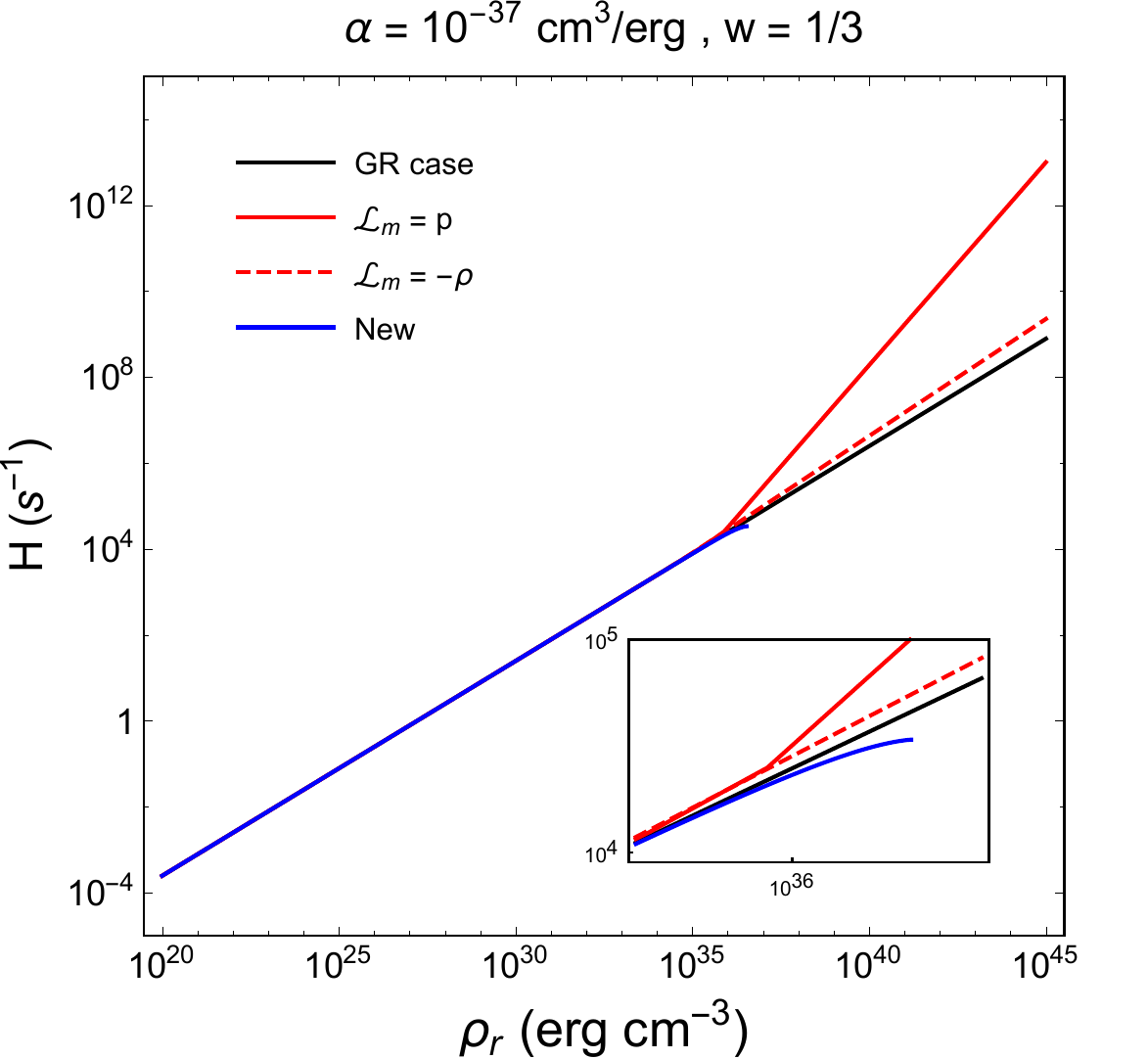}
	\includegraphics[width=0.45\textwidth]{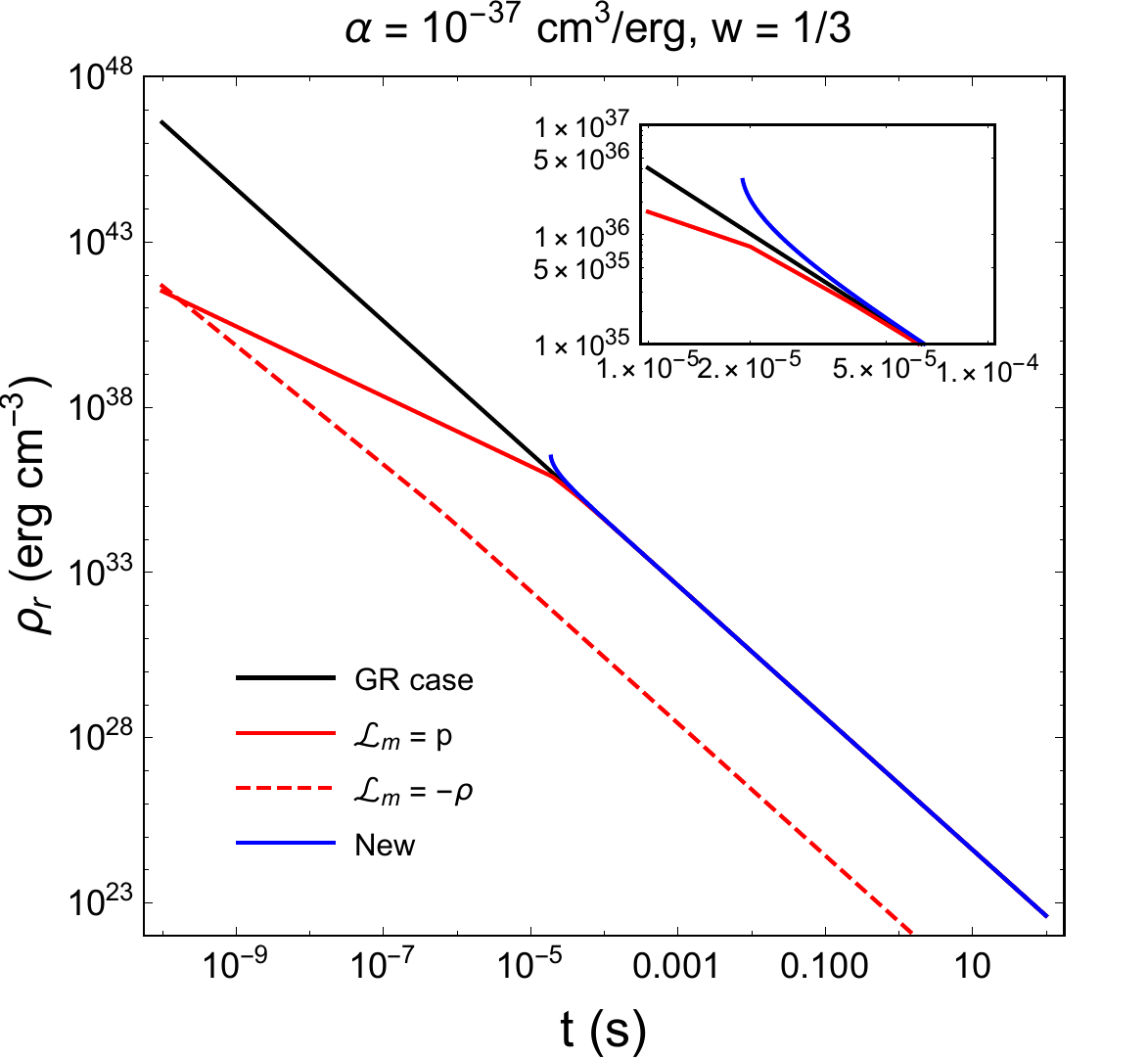}
	\caption{[Top] Hubble time parameter (H) against the energy density ($\rho$), and energy density against cosmic time ($t$) for dust ($w=0$) case. [Bottom] Same case, but for radiation ($w=1/3$) case}
	\label{fig:2}
\end{figure}

All profiles are shown in Fig.~\ref{fig:2}, where we now compare the discrepancies for each specific choice of $w$. Using the new formalism "New", the old prescription ($\mathcal{L}_m = p$, $\mathcal{L}_m = -\rho$) and the GR case with $\alpha = 10^{-37}~ \textrm{cm}^{3}/\textrm{erg}$ are plotted. For $\mathcal{L}_m = p$ for radiation dominated ($w=1/3$), the profiles are similar to \cite{Akarsu:2018zxl,Akarsu:2023agp}. The EMSG contribution leads to deviations from standard radiation in GR ($\alpha \rightarrow 0$), where $\rho_r = 3/32\pi t^2$. In EMSG, a quadratic equation arises for the radiation energy density; it is necessary to ensure that the choice of variables recovers GR as $\alpha$ vanishes. Analysis of the $w=1/3$ figure shows that, in the old prescription, $a \rightarrow \infty$ as $\rho_r \rightarrow 0$ for $t \rightarrow \infty$, and $a \rightarrow 0$ as $\rho_r \rightarrow \infty$ when cosmic time approaches zero ("big bang"). In the new prescription, Friedmann equation \eqref{friedman}--\eqref{cons} differs in sign and coefficient from the EMSG result in \eqref{friedmannrmsg}--\eqref{consemsg} and implies a bound at $\rho_r = 3/4\alpha$ for $\alpha>0,$ whereas reference \cite{Akarsu:2018zxl,Akarsu:2023agp} (for $\mathcal{L}_{m}=p$ case) gives $\rho_{r,max}=-1/8\pi$ for $\alpha<0$. In the early universe, energy density reaches a finite maximum within $10^{-5}~\textrm{s} \lesssim t \lesssim 10^{-4}~\textrm{s}$. To conclude, the figures indicate cosmological discrepancies, yet the new formalism remains close to GR for $\alpha = 10^{-37}~ \textrm{cm}^{3}/\textrm{erg}$.

\section{Conclusion}
\label{section6}

The standard formulation, in which the four-velocity $u_{\mu}$ and the metric tensor $g_{\mu \nu}$ are treated as independent variables, leads to nonuniqueness in the matter Lagrangian. By adopting the formalism of \cite{Akarsu:2023lre} for general curvature--matter coupling models, this study demonstrates that properly addressing the normalization condition of the four-velocity resolves this issue. Consequently, the framework produces models that are independent of the matter Lagrangian $\mathcal{L}_{m}$, regardless of whether $p$ or $-\rho$ is used. These results are applied to quark and neutron stars within the $f(R, T) = R+2\chi T$ gravity model with $\chi=1$ as an illustrative case. Discrepancies are observed in the choice of $\mathcal{L}_m$ between the previous and revised prescriptions: the quark star profile remains similar to general relativity (GR), whereas the maximum mass in the neutron star profile is higher than in GR. In the cosmological context, we explored the matter-dominated or dust ($w=0$), radiation-dominated ($w=1/3$), and the vacuum energy case ($w=-1$). In the radiation-dominated context, the EMSG model from \cite{Akarsu:2018zxl} is examined by analyzing its Friedmann and acceleration equations with $\alpha = 10^{-37}~ \textrm{cm}^{3}/\textrm{erg}$. The findings indicate that, in the early universe, the new prescription exhibits distinct behavior, including a finite maximum value during $10^{-5}~\textrm{s} \lesssim t \lesssim 10^{-4}~\textrm{s}$. Furthermore, the results suggest that the new formalism closely aligns with GR, despite existing discrepancies.

In conclusion, a universal matter Lagrangian has been identified that preserves the fundamental features of general relativity (GR). This result may provide a consistent framework for broader applications in modified gravity. The approach is particularly suitable for describing matter under extreme conditions, such as those relevant to the maximum mass of neutron stars, and for formulating dynamical Friedmann equations in infrared-scale cosmology.

%===========================================================

\acknowledgments
We thank Muhammad Fahmi Fauzi and Faris Ramadhantyo Darmawan for the useful discussions. In this work, we are supported by Hibah Fundamental DIKTI, PKS-178/UN2.RST/HKP.05.00/2026. \\

\appendix

\section{Variation of the normalization condition of 4-velocity}
\label{appA}
In this part, we derived a detailed variation with respect to metric tensor \cite{Akarsu:2023lre}. The starting line is as follows
\begin{eqnarray}
g^{\alpha\beta} u_{\alpha}u_{\beta}&=&-1, \\
\delta(g^{\alpha\beta} u_{\alpha}u_{\beta})&=&0, \nonumber \\ \frac{\delta g^{\alpha\beta}}{\delta g^{\mu\nu}}u_{\alpha}u_{\beta} + g^{\alpha\beta} \frac{\delta(u_{\alpha}u_{\beta})}{\delta g^{\mu\nu}} &=& 0, \nonumber \\ \frac{1}{2}(\delta^{\alpha}_{\mu}\delta^{\beta}_{\nu}+\delta^{\alpha}_{\nu}\delta^{\beta}_{\mu})u_{\alpha}u_{\beta} + g^{\alpha\beta} \frac{\delta(u_{\alpha}u_{\beta})}{\delta g^{\mu\nu}} &=& 0, \nonumber \\ g^{\alpha\beta} \frac{\delta(u_{\alpha}u_{\beta})}{\delta g^{\mu\nu}} &=&-u_{\mu}u_{\nu}, \nonumber \\ \underbrace{g^{\alpha\beta}u_{\alpha}u_{\beta}}_{-1} \frac{\delta(u_{\alpha}u_{\beta})}{\delta g^{\mu\nu}} &=& - u_{\alpha}u_{\beta}u_{\mu}u_{\nu}, \nonumber \\ \frac{\delta(u_{\alpha}u_{\beta})}{\delta g^{\mu\nu}} &=& u_{\alpha}u_{\beta}u_{\mu}u_{\nu}.
\end{eqnarray}

\section{The universal choice of $\mathcal{L}_{m}$}
\label{appB}

In this part, we will use the equations \eqref{pronew}-\eqref{ro2new} to the definition presented in (\ref{Theta}), (\ref{theta}), (\ref{gamma}), and (\ref{Xi}) and proof it in detail to show that the both choices are the same. We recall the equation \eqref{Theta}
\begin{eqnarray}
\Theta_{\mu\nu} (\mathcal{L}_{m}=p) &=& -2T_{\mu\nu}+ g_{\mu\nu}p -2g^{\alpha\beta} \frac{\delta^2 p}{\delta g^{\alpha\beta}\delta g^{\mu\nu}}, \\ &=& -2T_{\mu\nu} +g_{\mu\nu} p + (\rho+p)u_{\mu}u_{\nu},\\ &=& -2T_{\mu\nu}+T_{\mu\nu},\\ &=& -T_{\mu\nu}.
\end{eqnarray}
and 
\begin{eqnarray}
\Theta_{\mu\nu} (\mathcal{L}_{m}=-\rho) &=& -2T_{\mu\nu} -g_{\mu\nu}\rho + 2g^{\alpha\beta}\frac{\delta^2 \rho}{\delta g^{\alpha\beta}\delta g^{\mu\nu}}, \\ &=& -2T_{\mu\nu} -g_{\mu\nu}\rho + (\rho+p)(g_{\mu\nu}+u_{\mu}u_{\nu}), \\ &=&-2T_{\mu\nu}+T_{\mu\nu},\\ &=& -T_{\mu\nu}.
\end{eqnarray}
Threfore, we can infer that $\Theta_{\mu\nu} (\mathcal{L}_{m}=p)=\Theta_{\mu\nu} (\mathcal{L}_{m}=-\rho)$. Next, we move to the equation \eqref{theta}
\begin{eqnarray}
\theta_{\mu\nu} (\mathcal{L}_{m}=p) &=& p\left(g_{\mu\nu}T -2T_{\mu\nu} \right) -T T_{\mu\nu}+2T_{\mu}^{~\gamma}T_{\nu\gamma}  -4 T^{\alpha\beta} \frac{\delta^2 p }{\delta g^{\mu\nu}\delta g^{\alpha\beta}},
\end{eqnarray}
and 
\begin{eqnarray}
\theta_{\mu\nu} (\mathcal{L}_{m}=-\rho) &=& -\rho\left(g_{\mu\nu}T -2T_{\mu\nu} \right) -T T_{\mu\nu}+2T_{\mu}^{~\gamma}T_{\nu\gamma}  +4 T^{\alpha\beta} \frac{\delta^2 \rho }{\delta g^{\mu\nu}\delta g^{\alpha\beta}}, \\ &=& -\rho\left(g_{\mu\nu}T -2T_{\mu\nu} \right) -T T_{\mu\nu}+2T_{\mu}^{~\gamma}T_{\nu\gamma}  \nonumber \\ &&+ 4 T^{\alpha\beta} \bigg[ - \frac{\delta^2 p }{\delta g^{\mu\nu}\delta g^{\alpha\beta}} + \frac{(\rho+p)}{4}g_{\alpha\beta}g_{\mu\nu}-\frac{(\rho+p)}{2} g_{\mu\alpha} g_{\nu\beta} \bigg],\\ &=& 2T_{\mu}^{~\gamma}T_{\nu\gamma}-T T_{\mu\nu} -\rho\left(g_{\mu\nu}T -2T_{\mu\nu} \right)  -4T^{\alpha\beta} \frac{\delta^2 p }{\delta g^{\mu\nu}\delta g^{\alpha\beta}} \nonumber \\ &&+ g_{\mu\nu}T(\rho+p)-2T^{\alpha\beta}(\rho+p)g_{\mu\alpha}g_{\nu\beta}, \\ &=&  p\left(g_{\mu\nu}T -2T_{\mu\nu} \right) -T T_{\mu\nu}+2T_{\mu}^{~\gamma}T_{\nu\gamma}  -4 T^{\alpha\beta} \frac{\delta^2 p }{\delta g^{\mu\nu}\delta g^{\alpha\beta}} = \theta_{\mu\nu} (\mathcal{L}_{m}=p). \nonumber \\
\end{eqnarray}

Next, we shift to the equation \eqref{gamma}
\begin{eqnarray}
\Gamma_{\mu\nu}(\mathcal{L}_{m}=p) &=& -G_{\mu\nu}p -\frac{1}{2}RT_{\mu\nu} +2 R^{\alpha}_{\mu}T_{\alpha\nu}-2R^{\alpha\beta} \frac{\delta^2 p}{\delta g^{\mu\nu} \delta g^{\alpha\beta}},
\end{eqnarray}
and 
\begin{eqnarray}
\Gamma_{\mu\nu}(\mathcal{L}_{m}=-\rho) &=& G_{\mu\nu}\rho -\frac{1}{2}RT_{\mu\nu} +2 R^{\alpha}_{\mu}T_{\alpha\nu}+2R^{\alpha\beta} \frac{\delta^2 \rho}{\delta g^{\mu\nu} \delta g^{\alpha\beta}}, \\ &=& \left(R_{\mu\nu}-\frac{1}{2}g_{\mu\nu}R\right)\rho -\frac{1}{2}RT_{\mu\nu} +2 R^{\alpha}_{\mu}T_{\alpha\nu}\nonumber \\ && +2R^{\alpha\beta} \bigg[ - \frac{\delta^2 p }{\delta g^{\mu\nu}\delta g^{\alpha\beta}} + \frac{(\rho+p)}{4}g_{\alpha\beta}g_{\mu\nu}-\frac{(\rho+p)}{2} g_{\mu\alpha} g_{\nu\beta} \bigg],\\ &=& -G_{\mu\nu}p -\frac{1}{2}RT_{\mu\nu} +2 R^{\alpha}_{\mu}T_{\alpha\nu}-2R^{\alpha\beta} \frac{\delta^2 p}{\delta g^{\mu\nu} \delta g^{\alpha\beta}} =  \Gamma_{\mu\nu}(\mathcal{L}_{m}=p) .
\end{eqnarray}

Finally, the last gravity model we want to proof is $f(R,TG,TGD)$, where it has $\Xi_{\mu\nu}$ in $T_{\mu\nu}^{(\alpha)}$, and $\Theta_{\mu\nu}$ in $T_{\mu\nu}^{(\beta)}$. We only consider the $\Xi_{\mu\nu}$ here since the $\Theta_{\mu\nu}$ is the same expression as \eqref{Theta}
\begin{eqnarray}
\Xi_{\mu\nu} (\mathcal{L}_{m}=p) = -G_{\mu\nu}p +\frac{1}{2} G^{\alpha\beta}g_{\alpha\beta} (g_{\mu\nu}p-T_{\mu\nu})  -2 G^{\alpha\beta} \frac{\delta^2p}{\delta g^{\mu\nu} \delta g^{\alpha\beta}}.
\end{eqnarray}
For the $\mathcal{L}_{m}=-\rho,$ we have
\begin{eqnarray}
\Xi_{\mu\nu} (\mathcal{L}_{m}=-\rho) &=& G_{\mu\nu}\rho -\frac{1}{2} G^{\alpha\beta}g_{\alpha\beta} (g_{\mu\nu}\rho+T_{\mu\nu})  +2 G^{\alpha\beta} \frac{\delta^2\rho}{\delta g^{\mu\nu} \delta g^{\alpha\beta}}, \\ &=& G_{\mu\nu}\rho -\frac{1}{2} G^{\alpha\beta}g_{\alpha\beta} (g_{\mu\nu}\rho+T_{\mu\nu})  \nonumber \\&& +2 G^{\alpha\beta} \bigg[ - \frac{\delta^2 p }{\delta g^{\mu\nu}\delta g^{\alpha\beta}} + \frac{(\rho+p)}{4}g_{\alpha\beta}g_{\mu\nu}-\frac{(\rho+p)}{2} g_{\mu\alpha} g_{\nu\beta} \bigg], \\ &=&  -G_{\mu\nu}p +\frac{1}{2} G^{\alpha\beta}g_{\alpha\beta} (g_{\mu\nu}p-T_{\mu\nu})  -2 G^{\alpha\beta} \frac{\delta^2p}{\delta g^{\mu\nu} \delta g^{\alpha\beta}} = \Xi_{\mu\nu} (\mathcal{L}_{m}=p). \nonumber \\ 
\end{eqnarray}
Note that we go forward to the last line since the procedure is similar to the previous case that some terms involving $\rho$ are cancelled. All in all, the results are matched to what we have in \eqref{Thetanew}-\eqref{Xinew}.

%\newpage
%===========================================================

\end{document}